\documentclass[aps,prb,showpacs,twocolumn,floatfix]{revtex4}

\usepackage{graphics}

\def\tio2{TiO$_{\rm 2}$}
\def\o2{O$_{\rm 2}$}
\def\h2{H$_{\rm 2}$}
\def\coti{Co$_{\rm Ti}$}
\def\coint{Co$_{\rm int}$}
\def\vo{V$_{\rm O}$}
\def\cotivo{Co$_{\rm Ti}$V$_{\rm O}$}
\def\mub{$\mu_{B}$}
\def\tiint{Ti$_{\rm int}$}

\def\ef{$E_{F}$}

\begin{document}

\title{Theory of dopants and defects in Co-doped \tio2\ anatase}
\author{James M. Sullivan and Steven C. Erwin}
\affiliation{Center for Computational Materials Science,
Naval Research Laboratory, Washington, D.C. 20375}
\date{\today}

\begin{abstract}
We report first-principles microscopic calculations of the formation
energy, electrical activity, and magnetic moment of Co dopants and a
variety of native defects in \tio2\ anatase.  Using these results, we
use equilibrium thermodynamics to predict the resulting carrier
concentration, the average magnetic moment per Co, and the dominant
oxidation state of Co. The predicted values are in good agreement with
experiment under the assumption of O-poor growth conditions. In this
regime, a substantial fraction of Co dopants occupy interstitial sites
as donors. The incomplete compensation of these donors by
substitutional Co acceptors then leads to {\it n}-type behavior, as
observed experimentally.
\end{abstract}
\maketitle

\section{Introduction}\label{Introduction}
The recent discovery of room temperature ferromagnetism in Co-doped
\tio2\ anastase\cite{matsumoto01a} has led to a great deal of activity
both to understand the origins of ferromagnetic order in this material
and to raise the magnetic ordering temperature.
\cite{chambers01a,chambers02a,chambers02b,shinde02a,simpson02a,shim02a,soo02a,park02b} 
One promising line of inquiry is to understand first the role of
dopants and native defects in the \tio2\ host. For example, the
location of Co dopants in this material (substitutional versus
interstitial), their oxidation state, and their magnetic properties have been
the subject of intense scrutiny.
\cite{chambers01a,chambers02a,chambers02b,simpson02a,soo02a} Furthermore, it has been 
suggested that O vacancies, which are believed to give rise to the
observed {\it n}-type behavior in pure \tio2\ anatase,\cite{forro94a}
may provide free electrons which mediate the exchange interaction
between the Co dopants.\cite{chambers02a,chambers02b} This possibility
is very different from the hole mediated exchange interactions which
are believed to describe ferromagnetic order in a wide variety of
other dilute magnetic semiconductors, including InMnAs, GaMnAs, and
MnGe,\cite{park02a,ohno98a,dietl00a} and may have a direct bearing on
the origin of the anomalously high Curie temperature observed in
Co-doped \tio2\ anatase.

In this paper, we first use density-functional theory to determine the
electronic structure, formation energy, and electrical activity of Co
dopants and several native defects in \tio2\ anatase. (In the
remainder of this paper we will use ``defects'' to refer to both Co
dopants and native defects.) We then use standard methods to
calculate, as a function of temperature, the concentrations of each
defect in the \tio2\ host over a range of Co and O chemical potentials
relevant to different growth conditions. In summary, we find that O
vacancies do not play any significant role in Co-doped anatase.
Moreover, we find that the observed {\it n}-type behavior in Co-doped
\tio2\ strongly suggests that roughly half of the total Co content is in 
interstitial sites.  Under these conditions we find an
enhancement---relative to the Co$^{2+}$ low spin state---of the
average value of the local magnetic moment, in good agreement with
experiment.  Finally, under these conditions essentially all of the of
Co---both interstitial and substitutional---occurs in oxidation state
II, as observed experimentally.

\section{Background}\label{background}
We begin by briefly reviewing the relevant experimental results for
Co-doped \tio2\ anatase. We concentrate solely on experimental results
for samples in which Co is believed to be homogeneously distributed;
hence, we do not address Co clustering or its consequences.  We focus
on three observations which show good experimental reproducibility:
(1) the electrical nature of the samples (insulating versus {\it
n}-type or {\it p}-type), (2) the Co oxidation state, and (3) the
average magnetic moment per Co dopant.

Although the original work of Matsumoto and coworkers estimated the
ferromagnetic ordering temperature to be larger than 400 K, the average
magnetic moment per Co and {\it n}-type carrier concentration were
modest, 0.32 \mub\ and 10$^{18}$/cm$^{3}$ respectively. More recent efforts
have led to a larger magnetic moment per Co of 1.26
\mub\/,\cite{chambers01a,chambers02a,chambers02b} and to {\it n}-type 
carrier densities of 10$^{19}$/cm$^{3}$.\cite{chambers02c} These
increased values are
likely due to improvements in sample quality achievable with oxygen
plasma-assisted molecular-beam epitaxy (OPMBE).\cite{chambers00a} For example,
these samples were well characterized to rule out Co inclusions
as a source of the ferromagnetism.\cite{chambers02b,chambers02c} Using
both Co 2{\it p} photoemission and Co K-shell x-ray absorption
near-edge structure (XANES), the Co dopants in \tio2\ anatase were
shown to have a formal oxidation state of II. A strong
correlation between the magnetic and transport properties was
demonstrated: both highly doped and highly resistive samples are
typically non-magnetic, consistent with a picture of carrier-mediated
ferromagnetism competing with an antiferromagnetic superexchange
interaction.\cite{chambers02b,chambers02c}

\section{Theory}\label{Theoretical Approach}
\subsection{Formalism}
To understand the role of defects in \tio2\ anatase, we initially adopt
a simple picture of isolated impurities. In this approach, we first
determine the energy required for the defect to form in a given charge
state. We then use a standard thermodynamic approach to determine the
expected concentration of such defects at a given temperature. We 
assume that the defects do not interact, except
indirectly via charge transfer between them. Thus, for example, we do
not address the origin of the apparent ferromagnetic coupling between Co
dopants.

The defects we consider in this work are: interstitial Co (\coint\/),
substitutional Co on the Ti site (\coti\/), O vacancies (\vo\/),
interstitial Ti (\tiint\/), and defect complexes consisting of
nearest-neighbor pairs of substitutional Co and O vacancies
(\cotivo\/). There are two crystallographically distinct types of such
complexes: a \cotivo\ pair oriented along the $c$-axis and a \cotivo\
pair oriented nearly in the $ab$ plane.  We refer to these 
as \cotivo\/-{\it c} and \cotivo\/-{\it ab},
respectively.

We make no assumptions about which defects are donors and
which are acceptors, whether they are neutral or charged, or how many
are actually present. At a given temperature, the concentration of each
defect (in a given charge state) will be determined by its formation
energy. We use density-functional theory in a supercell approach to
calculate these formation energies according to 
\begin{equation}
E^q_{\rm form} = E^q_{t} - \sum_{j} n_{j}\mu_{j} + qE_{F}, \label{formenergy_eq}
\end{equation}
where $E^q_{t}$ is the total energy of a supercell containing one
defect in charge state $q$; $n_{j}$ and $\mu_{j}$ are the number and
chemical potential (energy per atom in the ground-state phase) of each
atomic species in the supercell; and \ef\ is the Fermi energy,
measured with respect to the valence band maximum (VBM) of the
host.\cite{singh85a,zhang98a,vandewalle00a,kohan00a,mahadevan02a}
Although the Fermi energy appears to be an independent variable here,
it is in fact determined by the constraint of electroneutrality, as
described below. Also, we note that although the total energies on the
right-hand side of Eq.~\ref{formenergy_eq} depend on whether
all-electron or pseudopotential methods are used, the formation energy
itself is analogous to a binding energy, and can therefore be
accurately calculated within either method.

In thermal equilibrium, the concentration of each defect, $D$, is determined 
by its formation energy:
\begin{equation}
C^{q}_{D} = N_{\rm sites}\exp(-E^q_{\rm form}/k_{B}T), \label{concentration_eq}
\end{equation}
where $N_{\rm sites}$ is the number of sites per unit volume available
to the defect.  Since the formation energies depend on the chemical
potentials and the Fermi level, it is evident that the concentrations
also depend on these quantities. The concentrations must also
satisfy the constraint of overall electrical neutrality; this provides
an additional equation that we use to determine \ef\ for any given
choice of chemical potentials. The electroneutrality condition must
take into account the contributions not only from charged defects, but
also from $p$ and $n$ (the hole and electron carrier densities). Thus,
for each given choice of the oxygen chemical potential, $\mu_{\rm O}$,
and cobalt chemical potential, $\mu_{\rm Co}$, the following equation
must be numerically solved:
\begin{equation}
p(E_{F}) - n(E_{F}) + \sum_{D,q}qC^{q}_{D}(E_F;\mu_{\rm O},\mu_{\rm Co}) = 0. \label{neutral_eq}
\end{equation}
Here, the sum is over all defects, $D$, in all possible charge states,
$q$.  The Ti chemical potential does not enter explicitly into this
equation for reasons discussed in the following section, and, of
course, the Co chemical potential is only relevant for
defects involving Co. The carrier densities {\it p} and {\it n}
are evaluated using the conventional
semiconductor expressions along with the {\it ab initio} density of states
of the host material, with a scissors operator applied to give the
experimental band gap. 

\subsection{Chemical Potentials}
The atomic chemical potentials, $\mu_{j}$, on the right hand side of
Eq.~\ref{formenergy_eq} are closely related to the experimental growth
conditions. A high value of chemical potential of a particular atomic
species is equivalent to a growth environment that is rich in that
species (in the sense of high partial pressure).

The chemical potentials of Ti and O which enter into Eq.~\ref{formenergy_eq} are not 
independent: equilibrium between the Ti and O atomic reservoirs and 
bulk \tio2\ anatase requires that
\begin{equation}
\mu_{\rm Ti}+\mu_{\rm O_{\rm 2}} = \mu_{\rm TiO_{\rm 2}}, \label{chempot_eq}
\end{equation}
where $\mu_{\rm TiO_{\rm 2}}$ is the total energy of bulk anatase. Moreover, 
in order to preclude the precipitation of bulk hcp Ti or O$_{\rm 2}$ dimers there
are additional thermodynamic restrictions on the individual chemical potentials:
\begin{equation}
\mu_{\rm O} < \mu_{\rm O}^{o}, \label{boundsOchempot_eq}
\end{equation}
and
\begin{equation}
\mu_{\rm Ti} < \mu_{\rm Ti}^{\rm bulk}, \label{boundTischempot_eq}
\end{equation}
where $\mu_{\rm O}^{o}$ is the energy per atom in an O$_{\rm 2}$ dimer
in its spin triplet ground state and $\mu_{\rm Ti}^{\rm bulk}$ is the
energy per atom in hcp Ti. It is conventional to refer to the upper
limit in Eq.~\ref{boundsOchempot_eq} as the O-rich limit and,
similarly, to the upper limit in Eq.~\ref{boundTischempot_eq} as the
Ti-rich limit. Because of the relationship between $\mu_{\rm Ti}$ and
$\mu_{\rm O}$ in Eq.~\ref{chempot_eq}, there are only two independent
variables in this approach: the O and Co chemical potentials. This
explains why the Ti chemical potential does not enter explicitly into
Eq.~\ref{neutral_eq}.

Thus, for a given choice of $\mu_{\rm Co}$ and $\mu_{\rm O}$, all the
other quantities (including $E_F$) in Eq.~\ref{neutral_eq} are
completely determined.  In practice, we eliminate $\mu_{\rm Co}$ as an
independent variable by constraining the total Co concentration to a
typical experimental value (5\%). Hence, in our formulation, the
concentration of all defects is entirely determined by the choice of O
chemical potential and the constraint of total Co concentration.

\subsection{Temperature}
Although temperature appears in both Eq.~\ref{formenergy_eq}
(implicitly) and Eq.~\ref{concentration_eq} (explicitly), our results
are not especially sensitive to this variable. Moreover, we stress
that the temperature appearing in Eq.~\ref{concentration_eq} has no
connection to the magnetic ordering temperature, and should instead be
understood simply as the temperature at which we evaluate the
concentration of defects in equilibrium with their respective
reservoirs. Upon completion of growth, we consider the reservoirs to
be disconnected and hence the number of these constituents to be
fixed. Thus we set the temperature to a particular value and examine
the behavior of the system as a function of the O chemical
potential. To this end, we calculate the defect concentrations using a
typical growth temperature of 873
K.\cite{matsumoto01a,chambers01a,shim02a} 

\subsection{Computational Details}
The total energy calculations were performed in a supercell consisting
of a 3$\times$3$\times$2 periodic repetition of the primitive unit
cell; thus, for the pure \tio2\ host, these supercells contain 108
atoms.  Structural relaxation was performed within a sphere of radius
of 4 \AA\ centered on the defect in question; calculations using a
sphere radius of 5 \AA\ give the same total energy to within 50
meV. This approach has proven both efficient and accurate for other
similar metal-oxide insulators.\cite{kohan00a} The total energy
calculations were evaluated in the local-density approximation (LDA)
within the ultrasoft pseudopotential formalism\cite{vanderbilt90a} as
implemented in the VASP code,\cite{kresse96a} with the zone center
used to sample the Brillouin zone of the supercells. A kinetic-energy
cutoff of 400 eV was used in all total energy evaluations. Only
defects involving Co atoms were treated in a spin-polarized
fashion. The ground state of bulk hcp Ti was treated non-magnetically,
bulk hcp Co was treated in an ferromagnetic configuration, and
the total energy of the \o2\ reference dimer was evaluated for the
spin triplet.

Since the Fermi energy in Eq.~\ref{formenergy_eq} is measured with
respect to the VBM of the host, we must align the VBM in the charged
defect supercell with that of the host material. To this end, we used
a local site average of the electrostatic potential to define a
reference energy; this local average was evaluated by a test charge
approach in which we calculate the electrostatic energy of a narrow Gaussian
charge distribution far from the defect.  We have checked that
using, as an alternative, the Ti 3{\it p} semi-core eigenvalues gives
very similar results. Including the zone sampling, kinetic-energy
cutoff, lattice relaxation and choice of reference energy, we estimate
the numerical uncertainty in our results to be 100--150 meV,
sufficient for addressing trends with respect to growth conditions.

To compute the total energy of a charged periodic system, we use a
standard procedure which systematically corrects for the artificial
and slowly converging Coulomb interaction between charged
defects.\cite{makov95a,kantorovich99a} A neutralizing homogeneous
background charge density is first added to the total charge density;
this makes the total energy a well-defined quantity. Next, the
artificial interaction among the charged defects within this
neutralizing background is subtracted from the total energy; this
interaction is estimated by expanding the defect-induced electron
density in a multipole series up to quadrupole order and then computing
the contribution to the total energy from the interaction of these
multipoles.\cite{quadnote}

Finally, we note that although the electronic band gap does not enter
explicitly into Eq.~\ref{formenergy_eq}, the LDA underestimation of
the gap does affect the formation energies of shallow donors, such as
O vacancies and interstitial Co. We have observed, however, that the
depths of these donor levels (relative to the conduction-band edge)
are essentially independent of the value
of the band gap. We demonstrated this by calculating the position of
the donor level using an artificially reduced lattice constant; as the
band gap increases with decreasing lattice constant, the donor level
closely tracks the conduction-band edge.  Thus, we have corrected the
formation energies of these donors using the experimental value of the
band gap and the depth of the donor levels given within LDA. In
addition, we assume that the LDA correctly predicts the formation
energies of the $+$2 charge states of these defects, because all
defect levels are empty in this charge state.

\section{Results and Discussion}\label{Results}
\subsection{Electronic Structure of \coti\/}\label{Elek_Struk}

\begin{figure}
\resizebox{8cm}{!}{\includegraphics{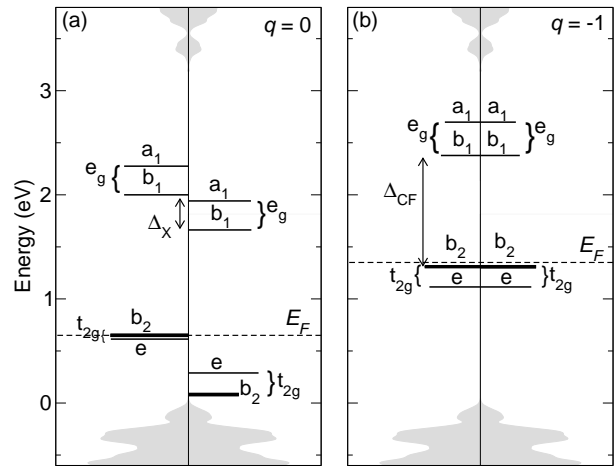}}%
\caption{\label{coti_n0dos_fig} First-principles energy-level diagram for
substitutional \coti\ in \tio2\ anatase. (a) Neutral charge state; (b)
Singly charged state, $q=-1$. The density of states of the anatase host is
shown as the gray shaded regions with a scissors operator applied to give the
experimental band gap. The length of each impurity level is proportional
to its {\it d} character. The exchange splitting, $\Delta_{X}$, and crystal field splitting,
$\Delta_{CF}$, are discussed in Sec. 4(c) (see text).}
\end{figure}

We discuss first the electronic structure of the isolated substitutional Co
dopant. Fig.~\ref{coti_n0dos_fig}(a) shows the single-particle levels
of neutral substitutional Co, as determined from the eigenvalue
spectra at $k=\Gamma$ using a 108-atom supercell.  The point group for
this defect is $D_{2d}$, so that all but one of the 3$d$-orbital
degeneracies are lifted by the crystal field; nevertheless, the
environment surrounding the Co dopant is still very nearly cubic, so
that the $e_g$ and $t_{2g}$ parentage of these levels is easily seen.
There are 3 occupied majority-spin levels and 2 occupied minority-spin
levels, yielding a net magnetic moment of 1 \mub\/.

It is instructive to study separately the exchange splitting and
crystal-field splitting. This separation occurs naturally in the
$q=-1$ charge state, for which the exchange splitting vanishes, as
shown in Fig.~\ref{coti_n0dos_fig}(b). In this case the largest
crystal-field splitting, about 1 eV, is between the $e_g$ and
$t_{2g}$ manifolds, with the $e_g$ manifold near the
LDA conduction-band edge and the $t_{2g}$ near the valence-band edge. The
$D_{2d}$ crystal field further splits the $e_g$ manifold into an upper
$a_1$ level of $z^2$ symmetry and a lower $b_1$ level of $x^2-y^2$
symmetry; this splitting is 0.3 eV. Likewise, the crystal field splits the
$t_{2g}$ manifold into an upper $b_2$ level of $xy$ symmetry and a
two-fold $e$ level of ($xz$, $yz$) symmetry; this splitting is 0.2 eV.

For charge states with non-zero magnetic moments, we find that the
exchange splitting varies approximately linearly with the magnetic
moment. (Since the orbital moment is strongly quenched,\cite{sullivan02a}
we regard the magnetic and spin moments as equivalent.)  For neutral
substitutional Co, the exchange splitting within the $e_g$ manifold is
0.2 eV for both levels. Within the $t_{2g}$ manifold, the exchange
splitting is strongly orbital-dependent: 0.5 eV for the $b_2$ levels
and 0.3 eV for the $e$ levels.

The calculated magnetic moments of substitutional Co in different
charge states, summarized in Table~\ref{tab_mag}, can now be easily
understood from the results of Fig.~\ref{coti_n0dos_fig}.  For the
neutral substitutional, there is a two-fold degenerate half-filled
minority spin level at the Fermi level. Removing an electron from this
level leads to a $+$1 charge state with a moment of 2
\mub\/. Likewise, adding an electron to this level leads to 
a $-$1 charge state with zero moment, since the $t_{2g}$
manifold is now completely filled.

The stable charge states and magnetic moments for interstitial Co are
quite different from the substitutional case. For example, we find
that all charge states experience off-center structural relaxations of 
order 1 \AA, lifting all remaining orbital degeneracies. The $+$2 charge 
state of interstitial Co, which is the lowest-energy charge state over most of
the gap, has a moment of 1 \mub\/. The $+$1 charge state
leads to a moment of 2 \mub\/, whereas the neutral configuration has a
moment of 1 \mub\/.

\subsection{Oxidation State}\label{oxid_state}
Since the correspondence between ``formal oxidation state'' and
``charge state'' will be an important issue here, we briefly summarize
the relationship between the two. A similar discussion has been given
for transition-metal impurities in GaP.\cite{singh85a} When a neutral
Co substitutes for Ti in \tio2, it takes on the oxidation state IV, the same
as that of Ti. Hence a substitutional \coti\ in the $-$1
charge state has a formal oxidation state of III, and a substitutional
\coti\ with charge $-$2 has an oxidation state of II. In a similar fashion
one can relate the oxidation state and charge state of interstitial
Co; in this case however, the charge state of the dopant is the same
as the oxidation state. Hence, neutral interstitial Co has oxidation
state 0, the charge state $+$1 has oxidation state I, and so
forth. The oxidation state of Co-related dopants in their various
stable charge states are summarized in Table~\ref{tab_mag}.

\begin{table}
\caption{\label{tab_mag}
Stable charge states of various Co dopants in \tio2\ anatase, with
their formal oxidation states and calculated magnetic moments.  The
results for the $-$2 charge state of \coti\ are obtained from a model
described in the text. The oxidation state values in parentheses for
\cotivo\ complexes are based on the assumption that the neutral
complex can be represented as \coti\ and \vo\ in the $-$2 and $+$2
charge states, respectively.}
\begin{tabular}{lccc}
\toprule
Defect      & Charge & Oxidation state & $M$ ($\mu_{B}$) \\
\colrule
\coti              & 1   &   V      & 2.0   \\
                   & 0   &  IV      & 1.0   \\
                   & $-$1   &  III     &  0.0   \\
                   & $-$2   &  II      &  1.0   \\
\colrule
\coint\            & 2   &  II      &  1.0   \\
                   & 1   &   I      &  2.0   \\
                   & 0   &  0       &  1.0   \\
\colrule
\cotivo\ -{\it c}  & 1   &  --     &  2.0   \\
                   & 0   & (II)      &  1.0   \\
\\
\cotivo\ -{\it ab} & 1   &  --     &  0.0   \\
                   & 0   & (II)      &  1.0   \\
\botrule
\end{tabular} 
\end{table}

\subsection{The $-$2 Charge State of \coti\/}\label{lda_coti}
In the LDA calculation there is no stable $-$2 charge state of
substitutional Co: upon adding an additional electron to the stable
$-$1 charge state, we find no dopant-derived level within the gap, but
rather partial occupation of the LDA conduction bands.  This is
problematic, since the experimental finding of an oxidation state of
II for Co would be consistent with a $-$2 charge state for
substitutional Co. Here we investigate whether the absence of a $-$2
charge state within LDA is due to the well-known underestimation of
the band gap by the LDA (the LDA predicts a band gap of 2.2
eV compared to the experimental value of 3.2 eV\cite{tang77a}). 

Rather than attempting to correct the LDA band gap, we propose a more 
direct description and try to estimate the formation energy of the
$-$2 charge state using our results for the energy levels of the $-$1
charge state. Since the formation energy is in general a linear
function of the Fermi energy, we do this in terms of the ``energy of
transition,'' $E(-/--)$, which is defined as that value of $E_F$ for
which $E_{\rm form}^{q=-1}$ is equal to $E_{\rm form}^{q=-2}$.
Referring to Fig.~\ref{coti_n0dos_fig}(b), we model the $-$2 charge
state by occupying the first available impurity level in the $q=-1$
level diagram, namely the empty $b_1$ level.  In this diagram,
the $b_1$ level is at $E_F+\Delta_{CF}$, so that occupying it will shift
the Fermi level upward by an amount $\Delta_{CF}$. However, by singly
occupying this level, one expects an accompanying exchange
splitting and thus a downward shift of the Fermi level.  
Hence, we estimate the change in the Fermi
energy between the $-$1 and $-$2 charge state to be given by
\begin{equation}
E(-/--) = E(0/-) + \Delta_{CF} - \Delta_{X}, \label{e2_eqtn}
\end{equation}
where $E(0/-)$ is the energy of transition between the neutral and
$-$1 charge states; $\Delta_{CF}$ is the crystal-field
splitting between the $b_{2}$ and $b_{1}$ levels; and $\Delta_{X}$ is
the exchange splitting due to the unpaired $b_{1}$ electron.

We approximate $\Delta_{X}$ by the $b_1$ exchange splitting for the neutral state;
this is 0.3 eV, as shown in Fig.\ref{coti_n0dos_fig}(a).  We neglect
the onsite energy which arises from the interaction of the added
$b_{1}$ electron with the $t_{2g}$ electrons, but we expect this
energy to be significantly smaller than the crystal-field energy, and
hence expect Eq.~\ref{e2_eqtn} to be reasonably accurate.

A similar proposal could be made for more-negative charge states, for
example, the $-$3 charge state of substitutional Co. However, we
should then include an additional term, $U$, for the onsite
interaction between two $b_{1}$ electrons of opposite spin. This
onsite energy will be large and can be estimated as
\begin{equation}
U \geq E(0/-) - E(+/0) - \Delta_{X}, \label{U_eqtn}
\end{equation}
where $\Delta_{X}$ = 0.3 eV is again taken from the neutral
configuration, but in this case accounts for the exchange energy lost
when the moment is reduced from 1 \mub\ to zero in the transition from
the neutral to the $-$1 charge state.\cite{umodel}
We find $U \geq$ 1.0 eV and, hence, the $-$3 charge state of
substitutional Co will lie well above the conduction-band edge, so
that it (and more highly charged negative states) can be
excluded from further consideration.

\subsection{Formation Energies in the O-rich Limit}\label{orich_limit}
\begin{figure}
\resizebox{8cm}{!}{\includegraphics{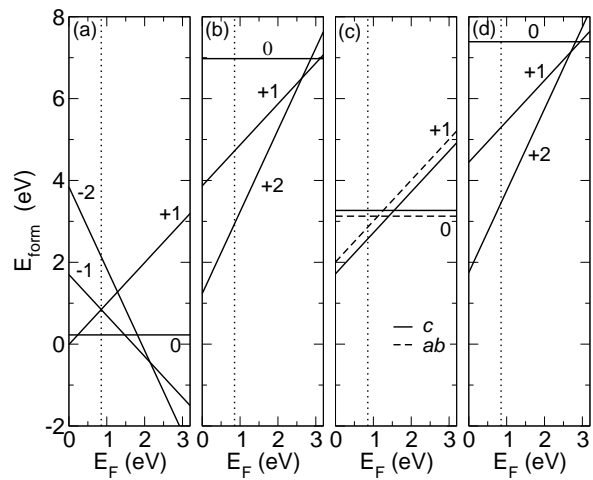}}%
\caption{\label{orich_co0.05_fig} Formation energies of (a) \coti, (b) \coint\/, 
(c) \cotivo\/, (d) \vo\ defects as a function of Fermi level in the
O-rich limit. The Co chemical potential was chosen to give a total Co
concentration of 5\%\/. The Fermi energy which satisfies
electroneutrality for these choice of growth conditions and Co
concentration is denoted by the dotted vertical line. The stable
charge states for each defect are labeled.}
\end{figure}

To address the characteristics of Co-doped samples grown with OPMBE
we first consider conditions O-rich growth conditions.
Figure~\ref{orich_co0.05_fig} shows the defect formation energies
versus Fermi energy in the O-rich limit, with the total Co
concentration constrained to be 5\%\/. In comparison to experiment
this scenario has three main shortcomings: (1) the average magnetic
moment is 1 \mub\/, significantly smaller than the measured
value\cite{chambers02a,chambers02b} of 1.26 \mub\/; (2) the Fermi
level is well below midgap, whereas experimental measurements show the
material to be {\it n}-type;\cite{matsumoto01a,chambers02b,chambers02c} (3) Co appears
almost entirely as neutral substitutionals (oxidation state IV),
whereas Co 2{\it p} photoemission and XANES suggest the oxidation
state of Co is II.\cite{chambers02b,chambers02c} These discrepancies
suggest that the conditions present during growth of the samples
are not O-rich.

\subsection{Variation with O Chemical Potential}\label{ochempot_vary}
\begin{figure}
\resizebox{8cm}{!}{\includegraphics{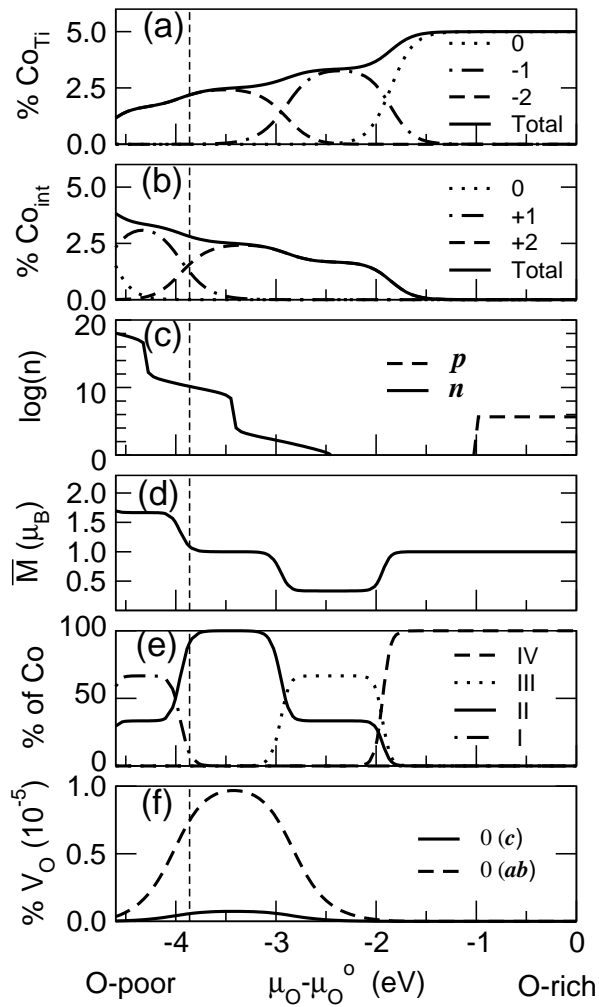}}%
\caption{\label{obs_vs_O_0_05_fig} Variation, versus O chemical
potential, of (a) \coti\ concentration, 
(b) \coint\ concentration, (c) carrier concentration, (d) average
magnetic moment per Co, (e) percent of Co in various oxidation states,
and (f) concentration of \vo\/-related defects versus O chemical
potential. In panels (a) and (b) the concentration of each charge
state of Co is indicated as well as the total concentration. In panel
(f) only the concentration of \cotivo\ defects are shown, since the
concentration of isolated O vacancies is much smaller. Values of the
O chemical potential to the left of the dashed vertical line are those
for which our theoretical results agree qualitatively with experiment 
(see text).}
\end{figure}

To determine what growth conditions give rise to the observed magnetic 
and transport properties of Co-doped anatase samples, we study the 
consequences of varying the O chemical potential away from 
its upper limit. In Fig.~\ref{obs_vs_O_0_05_fig} we plot, versus the O chemical potential,
the following quantities: concentration of of substitutional and
interstitial Co; {\it n} and {\it p} carrier densities; average
magnetic moment per Co; percentage of Co in different oxidation
states; and concentration of O vacancy-related defects (the
concentration of interstitial Ti is negligible throughout this range of O
chemical potential). For every value of the O chemical potential, the
total concentration of Co was constrained to 5\%\/.

\subsubsection{Cobalt concentration}

For values of the O chemical potential within $\sim$1.5 eV of the
O-rich limit, the substitutional site remains the preferred Co
site.  The situation is very different in the O-poor limit. In this
regime, interstitial Co (in the $+$1 and $+2$ charge states) becomes
energetically competitive with substitutional Co, and hence the
concentrations of substitutional and interstitial Co become
comparable. This results from a subtle balance of the O and Co
chemical potentials. As one decreases the O chemical potential from
its upper limit, the formation energies of substitutional dopants
increases, since it costs more energy to replace Ti with Co as one
moves toward the Ti-rich limit.  Of course, as a function of the O
chemical potential, the Co chemical potential must also change,
increasing monotonically to maintain the fixed total Co
concentration. At $\sim$1 eV below the O-rich limit, interstitial Co
begins to play a role; the preference for the interstitial site
increases monotonically below this point, eventually accounting for
most of the total Co concentration in the O-poor limit.

\subsubsection{Carrier densities}
The defect concentrations shown in Fig.~\ref{obs_vs_O_0_05_fig} were
computed for a temperature of 873 K, as mentioned earlier. However,
since transport, magnetometry, and photoemission measurements are
generally performed at room temperature, we have used a more relevant
temperature of 300 K to compute the carrier concentrations, magnetic
moments, and oxidation-state fractions---while keeping the total defect
concentrations themselves frozen at their high-temperature values.

Figure~\ref{obs_vs_O_0_05_fig}(c) shows the logarithm of the hole and
electron carrier densities evaluated at 300 K as a function of the O
chemical potential. Near the O-rich limit the carrier densities are
nearly constant, since the Fermi level maintains a value of $\sim$0.8
eV relative to the VBM due to compensation by equal but small numbers
of substitutional Co in the $-$1 and $+$1 charge states. In this
region the material is very weakly {\it p}-type with a hole
concentration $\sim$10$^{6}$/cm$^{3}$ and will likely appear
insulating in transport measurements. 

At $\sim$1 eV below the O-rich limit, the appearance of interstitial
Co leads to partial compensation of the Co substitutional defects.
This compensation drives the Fermi level towards the conduction-band
edge, and leads to a marked increase in the electron
concentration as the O chemical potential is further reduced.  The 
step-wise behavior of the carrier density as the O chemical potential is
decreased results from the Fermi level passing through donor levels
due to interstitial Co. At the O-poor limit, we find $n\sim 10^{20}$/cm$^3$.  Thus, we suggest
that the experimentally observed {\it n}-type behavior results from the
incomplete compensation of interstitial Co by substitutional Co, which
facilitates the thermal excitation of electrons from interstitial Co into the
conduction band.

\subsubsection{Magnetic moment}

In Fig.~\ref{obs_vs_O_0_05_fig}(d) we show the average magnetic moment per Co,
defined here as
\begin{equation}
\overline{M} = \sum_{D,q}M^{q}_{D}C^{q}_{D}/\sum_{D',q'}C^{q'}_{D'}, \label{magmom_eq}
\end{equation}
where $M^{q}_{D}$ is the magnetic moment of a Co-related defect in
charge state $q$.  We assume a ferromagnetic alignment of all the
moments, and thus $\overline{M}$ can be considered an upper bound for
the measured average value of the moment per Co.

The average moment shows a non-monotonic variation with O chemical
potential. In the O-rich region it is constant with a value of 1
\mub\/, because neutral substitutional Co is the dominant 
defect. For intermediate values of the O chemical potential,
$\overline{M}$ is less than 1 \mub\/ due to the presence of
substitutionals in the $-1$ charge state, which have zero moment. In
the O-poor limit the average moment per Co is larger than 1
\mub\/, due to the appearance of interstitial Co in the $+$1 
charge state, which has a moment of 2 \mub\/. Hence, only in the O-poor
limit do we obtain a moment per Co larger than the Co$^{2+}$
low-spin value of 1 \mub\/, and thus consistent with experiment.

\subsubsection{Oxidation states}

Figure~\ref{obs_vs_O_0_05_fig}(e) shows the percentage of Co dopants
in different oxidation states. In the O-rich limit the oxidation state
of Co is IV, since it occurs primarily as a neutral
substitutional. This oxidation state dominates until $\sim$1.5 eV
below the O-rich limit where oxidation state III, resulting from
substitutionals in the $-$1 charge state, is briefly dominant. Between
3 and 4 eV below the O-rich limit the predominant oxidation state of
Co is II, the same as that deduced from the 2{\it p} photoemission and
XANES results of Refs.\onlinecite{chambers02b}
and \onlinecite{chambers02c}. In the O-poor limit the oxidation state
is primarily I, since the dominant type of defect is interstitial Co in
the $+$1 charge state. Although this oxidation state has not been
observed experimentally, comparison to photoemission or XANES results
on samples with known Co oxidation state of I has not been
established.

\subsubsection{O vacancies}

For all thermodynamically allowed chemical potentials, the
concentration of isolated O vacancies is negligible, as well as that
of \cotivo\ complexes; their concentrations are shown in
Fig.~\ref{obs_vs_O_0_05_fig}(f) to be never above 10$^{-5}$
\%\/. These low concentrations result from the fact that the formation
energies of these two defects are much higher than the thermal energy
$k_{B}T = 60$ meV; this is clear from Fig.~\ref{orich_co0.05_fig},
where the formation energy of both isolated \vo\ and \cotivo\
complexes is greater than 2 eV for any value of the Fermi level. Thus
O vacancies, either isolated or in complexes with substitutional Co,
will play no significant role in determining the carrier concentration
in Co-doped samples.

We note that the formation energies of O vacancies in
Fig.~\ref{orich_co0.05_fig} and the deduced donor level positions in
the band gap are incompatible with the interpretation of the observed
{\it n}-type conductivity of pure \tio2\/ samples.\cite{forro94a} Even
in the O-poor limit, the LDA does not lead to a carrier density with
the observed value 10$^{18}$/cm$^{3}$, nor does it give the same
temperature dependence of the carrier concentration, as the energy for
activation is quite different: $\sim$200 meV (the position of the
highest O vacancy donor level below the conduction band edge) in the
LDA versus 4.2 meV in the results of Ref. \onlinecite{forro94a}. We do
not have a definitive resolution to this apparent disagreement; we
have checked, however, that if we adjust the O vacancy formation
energies to reproduce the experimentally observed carrier density and
temperature dependence in undoped samples, the conclusions of this
work are unchanged. 

Regarding the experimental interpretation that O
vacancies are also the source of {\it n}-type carriers in Co doped
samples,\cite{chambers02a,chambers02b} we note that the observed
average moment per Co of 1.26 \mub\/ can only be reproduced if some
fraction of Co interstitials in the $+$2 charge state are present, as
only these defects have a moment larger than 1 \mub\/. Neither
substitutional nor interstitial Co has any significant orbital
moment\cite{sullivan02a} so that the observed magnetic moment per Co
can only result from the statistical distribution of Co moments in
interstitial (2 \mub\/) and substitutional (1 \mub\/) sites. In such a
situation the Fermi level and carrier concentration are determined
solely by the Co dopants.

\subsubsection{Conclusion}

Considering these trends in the {\it n}-type carrier density, average
magnetic moment per Co, and oxidation state of Co, we suggest the
actual conditions of growth of Co-doped \tio2\ anatase are O-poor,
corresponding to O chemical potentials between 3.8 and 4.6 eV below
the O-rich limit (values of the O chemical potential to the left of 
the dashed vertical line in Fig.~\ref{obs_vs_O_0_05_fig}). Under these type of
conditions we obtain qualitatively and quantitatively good agreement
with experimental observations.

\section{Summary}\label{Summary}
In summary, we have examined the role of native defects and Co dopants in \tio2\ anatase over a
range of chemical potentials corresponding to different growth conditions. Under O-rich growth
conditions we find that Co dopants will be formed primarily in neutral substitutional form 
corresponding to oxidation state IV, an average magnetic moment of 1 \mub\, and insulating 
electrical character. These results are in conflict with the experimentally observed behavior
of Co-doped samples and suggest that the growth conditions are more likely to be O-poor.
O-poor conditions lead to roughly equal concentrations of substitutional and 
interstitial Co, {\it n}-type behavior resulting from thermal excitation of electrons from 
interstitial Co into the conduction band, and an average magnetic moment per Co in good agreement
with experiment.

\begin{acknowledgments}
One of the authors (J.M.S.) acknowledges the National Research
Council for support during this work in the form of a postdoctoral associateship. 
This work was  funded in part by DARPA and ONR. Computational work was supported 
in part by a grant of HPC time from the DoD Major Shared Resource Center ASCWP.
\end{acknowledgments}

\bibliographystyle{/home/main1/hellberg/bib/prsty}

\begin{thebibliography}{10}

\bibitem{matsumoto01a}
Y. Matsumoto, M. Murakami, T. Shono, T. Hasegawa, T. Fukumura, M. Kawasaki, P.
  Ahmet, T. Chikyow, S. Koshihara, and H. Koinuma, Science {\bf 291},  854
  (2001).

\bibitem{chambers01a}
S.~A. Chambers, S. Thevuthasan, R.~F.~C. Farrow, R.~F. Marks, J.~U. Thiele, L.
  Folks, M.~G. Samant, A.~J. Kellock, N. Ruzycki, D.~L. Ederer, and U. Diebold,
  Appl.\ Phys.\ Lett.\ {\bf 79},  3467  (2001).

\bibitem{chambers02a}
S.~A. Chambers, Materials Today {\bf April},  34  (2002).

\bibitem{chambers02b}
S.~A. Chambers, S.~M. Heald, R.~F.~C. Farrow, J.-U. Thiele, R.~F. Marks, M.~F.
  Toney, and A. Chattopadhyay, cond-mat/0208315  (2002).

\bibitem{shinde02a}
S.~R. Shinde, S.~B. Ogale, S. Das~Sarma, S.~E. Lofland, C. Lanci, J.~P. Buban,
  N.~D. Browning, V.~N. Kulkarni, J. Higgins, R.~P. Sharma, R.~L. Greene, and
  T. Venkatesan, cond-mat/0203576  (2002).

\bibitem{simpson02a}
J.~R. Simpson, H.~D. Drew, S.~R. Shinde, Y. Zhao, S.~B. Ogale, and T.
  Venkatesan, cond-mat/0205626  (2002).

\bibitem{shim02a}
I.-B. Shim, S.-Y. An, C.~S. Kim, S.-Y. Choi, and Y.~W. Park, J. Appl.\ Phys.\
  {\bf 91},  7914  (2002).

\bibitem{soo02a}
Y.~L. Soo, G. Kioseoglou, S. Kim, Y.~H. Kao, P. Sujatha~Devi, J. Parise, R.~J.
  Gambino, and P.~I. Gouma, Appl.\ Phys.\ Lett.\ {\bf 81},  655  (2002).

\bibitem{park02b}
M.~S. Park, S.~K. Kwon, and B.~I. Min, Phys.\ Rev.\ B {\bf 65},  161201(R)
  (2002).

\bibitem{forro94a}
L. Forro, O. Chauvet, D. Emin, L. Zuppiroli, H. Berger, and F. Levy, J. Appl.\
  Phys.\ {\bf 75},  633  (1994).

\bibitem{park02a}
Y.~D. Park, A.~T. Hanbicki, S.~C. Erwin, C.~S. Hellberg, J.~M. Sullivan, J.~E.
  Mattson, T.~F. Ambrose, A. Wilson, G. Spanos, and B.~T. Jonker, Science {\bf
  295},  651  (2002).

\bibitem{ohno98a}
H. Ohno, Science {\bf 281},  951  (1998).

\bibitem{dietl00a}
T. Dietl, H. Ohno, F. Matsukura, J. Cibert, and D. Ferrand, Science {\bf 287},
  1019  (2000).

\bibitem{chambers02c}
S. A. Chambers (private communication).

\bibitem{chambers00a}
S.~A. Chambers, Solid State Commun. {\bf 39},  105  (2000).

\bibitem{singh85a}
V.~A. Singh and A. Zunger, Phys.\ Rev.\ B {\bf 31},  3729  (1985).

\bibitem{zhang98a}
S.~B. Zhang, S.-H. Wei, A. Zunger, and H. Katayama-Yoshida, Phys.\ Rev.\ B {\bf
  57},  9642  (1998).

\bibitem{vandewalle00a}
C.~G. Van~de Walle, Phys.\ Rev.\ Lett.\ {\bf 85},  1012  (2000).

\bibitem{kohan00a}
A.~F. Kohan, G. Ceder, D. Morgan, and C.~G. Van~de Walle, Phys.\ Rev.\ B {\bf
  61},  15019  (2000).

\bibitem{mahadevan02a}
P. Mahadevan and A. Zunger, Phys.\ Rev.\ Lett.\ {\bf 88},  047205  (2002).

\bibitem{vanderbilt90a}
D. Vanderbilt, Phys.\ Rev.\ B {\bf 41},  7892  (1990).

\bibitem{kresse96a}
G. Kresse and J. F{\"u}rthmuller, Phys.\ Rev.\ B {\bf 54},  11169  (1996).

\bibitem{makov95a}
G. Makov and M.~C. Payne, Phys.\ Rev.\ B {\bf 51},  4014  (1995).

\bibitem{kantorovich99a}
L.~N. Kantorovich, Phys.\ Rev.\ B {\bf 60},  15476  (1999).

\bibitem{quadnote}
We assume that the monopole-quadrupole interaction energy in a tetragonal cell
  is of the same form as that for a simple cubic cell.\cite{makov95a} This
  assumption is confirmed by calculations of atomic ionization energies in
  tetragonal supercells of the same size and shape as that of our defect
  supercells.

\bibitem{sullivan02a}
J. M. Sullivan and S. C. Erwin (unpublished).

\bibitem{tang77a}
T. Tang, H. Berger, P.~E. Schmid, F. Levy, and G. Burri, Solid State Commun.
  {\bf 23},  161  (1977).

\bibitem{umodel}
We assume this is a lower bound, since the $e_{g}$ orbitals are more localized
  than the $t_{2g}$ orbitals, and thus can be expected to have a larger onsite
  interaction.

\end{thebibliography}

\end{document}